\documentclass[conference]{IEEEtran}
\IEEEoverridecommandlockouts
\usepackage{cite}
\usepackage{amsmath,amssymb,amsfonts}
\usepackage{algorithmic}

\usepackage{cuted}
\usepackage{multirow}
\usepackage{statmath}
\usepackage{longtable}
\usepackage[export]{adjustbox}
\usepackage{graphicx, subcaption, caption}
\usepackage[font=small]{caption}
\usepackage{titlesec}
\titlespacing*{\section}
{0pt}{6pt}{6pt}
\titlespacing*{\subsection}
{0pt}{6pt}{2pt}
\usepackage{lipsum}
\usepackage{hyperref}
\usepackage{textcomp}
\usepackage{xcolor}
\emergencystretch=\maxdimen
\usepackage{siunitx}
\usepackage[bottom]{footmisc}
\usepackage[utf8]{inputenc}
\usepackage[T1]{fontenc}
\usepackage{booktabs, makecell, ltablex, threeparttablex}

\emergencystretch=\maxdimen
\def\BibTeX{{\rm B\kern-.05em{\sc i\kern-.025em b}\kern-.08em
    T\kern-.1667em\lower.7ex\hbox{E}\kern-.125emX}}
\begin{document}

\title{{Towards Underwater Detection of Diluted Bitumen with Interdigital Sensors}

 \thanks{We acknowledge the support of the Natural Sciences and Engineering Research Council of Canada (NSERC) [funding reference number RGPIN-2021-04049] and the Ingenuity Labs Research Institute for partially funding this research.}
}

\author{
\IEEEauthorblockN{
Graziella Bedenik, Melissa Greeff, and Matthew Robertson }

\IEEEauthorblockA{Queen's University, Kingston, Canada \\
\{graziella.bedenik, melissa.greeff, m.robertson\}@queensu.ca } 

\vspace{-6ex}}

\maketitle

\begin{abstract}
Diluted bitumen spills in freshwater environments pose significant challenges for detection and remediation due to the complex behaviour of the material once released, rendering traditional oil monitoring techniques, such as aerial surveys and fluorometry, ineffective. This study investigates the feasibility of utilizing interdigital capacitive sensors to detect dilbit in water via frequency-dependent impedance measurements. Three sensor designs were fabricated and tested with a vector network analyzer across samples with varying dilbit concentrations.  All sensor designs detected electrical changes, exhibiting linear regions and strong potential for quantitative sensing. These findings support the integration of the sensors into autonomous underwater robotic platforms for real-time, distributed monitoring of dilbit spills in freshwater systems.
\end{abstract}


\section{Introduction}
\label{sec:intro}

Freshwater ecosystems are essential to biodiversity, public health, and regional economies. Canada, which holds the world’s largest surface freshwater reserves \cite{stoyanovich2022fate}, faces increasing threats to these systems from anthropogenic stressors, including municipal and industrial pollution \cite{rolsky2020municipal} and eutrophication-driven algal blooms \cite{binding2021eolakewatch}. Despite their importance, freshwater environments remain significantly under-researched compared to terrestrial and marine systems \cite{su2021human}.

Among the most complex freshwater pollutants in Canada is diluted bitumen (dilbit), a heavy crude oil product transported by rail and pipeline. Produced by blending viscous bitumen with diluents, dilbit becomes denser as the diluent evaporates post-spill, leaving behind adhesive residue that submerges or sinks, rendering detection, recovery, and containment difficult \cite{national2016spills}.

Conventional oil detection methods—such as aerial imaging, sonar, fluorometry, and diver observations—are limited in identifying submerged or weathered dilbit \cite{national2016spills,fingas2015diluted}. These approaches are costly, sensitive to environmental conditions, and lack continuous or distributed coverage. Even fluorometry, a widely used method, performs inconsistently when applied to weathered or sunken dilbit \cite{ji2024optimized}, prompting the need for robust, embeddable, real-time alternatives.

Subaquatic robots have emerged as promising platforms for environmental monitoring \cite{dunbabin2012robots,bogue2023role}, with growing interest from Canadian institutions \cite{lindsay2022collaboration,bedenik2025bistable}. However, suitable sensing technologies for freshwater integration remain a challenge. Electrochemical sensors have demonstrated pollutant detection capabilities in marine environments \cite{haghighi2025innovative}, and capacitive sensors have been used to estimate oil film thickness in surface spills \cite{saleh2018situ,saleh2022dual,saleh2024situ}. Still, these approaches are not directly optimized for underwater dilbit detection.

Although changes in water’s electrical properties can indicate hydrocarbon presence, instrumentation-specific aspects, particularly for dilbit, remain underexplored. Prior work has not addressed the frequency-dependent behaviour of electrical responses, nor assessed which quantities (e.g., conductance, capacitance, impedance) are most selective or practical for embedded sensing. Experimental studies of dilbit’s electrical signature, especially in robot-compatible formats, are notably lacking.

\begin{figure}[t]
    \centering
    \includegraphics[width=0.7\linewidth]{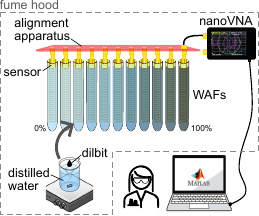}
    \caption{Condensed view of the experimental setup used in this study, highlighting the key components: interdigital sensors, WAF samples prepared from distilled water and dilbit, the VNA, a custom-manufactured sensor alignment apparatus, and the data acquisition laptop.}
    \label{fig:figure1}
\end{figure}

This study addresses these gaps by investigating interdigital capacitive sensors for underwater dilbit detection. Their planar geometry and compatibility with embedded systems make them well-suited for robotic integration. The sensors are designed to operate in high-frequency regimes, where capacitive effects dominate and electrolysis or noise are minimized. Water-accommodated fraction (WAF) samples with varying dilbit concentrations were analyzed using a vector network analyzer (VNA) to identify frequency ranges with maximum sensitivity. The experimental setup is shown in Fig.~\ref{fig:figure1}. The main contributions are: (1) frequency-dependent impedance analysis in dilbit-contaminated water; (2) identification of favourable operating frequencies; (3) evaluation of three sensor geometries; and (4) insights into integration with autonomous freshwater monitoring systems.


\section{Sensor Design and Fabrication}
\label{sec:sensor}

\begin{figure}[t]
    \centering
    \includegraphics[width=0.7\linewidth]{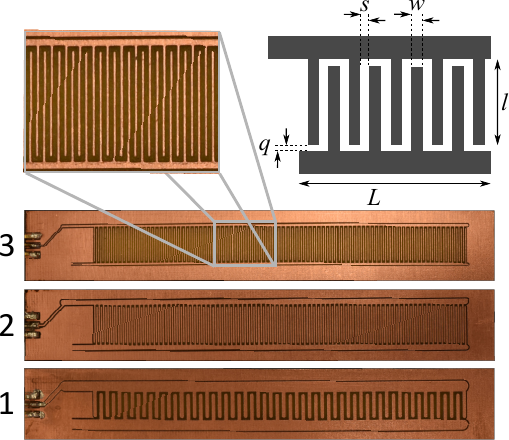}
    \caption{Manufactured interdigital sensors and their simplified design parameters. Adapted from \cite{teodorescu2024constraints}.}
    \label{fig:IDT}
\end{figure}

This study evaluates custom-designed interdigital capacitive sensors for detecting varying concentrations of dilbit WAFs in distilled water. These sensors feature an interdigitated electrode structure—two sets of parallel electrodes arranged in an alternating planar layout, as shown in Fig.~\ref{fig:IDT}. Depending on the operating frequency, each digit interacts differently with the surrounding medium, affecting the sensor's impedance response \cite{teodorescu2024constraints}. The sensing principle relies on measuring impedance variations induced by changes in material properties, such as the presence and concentration of a target chemical.

Dilbit contamination alters the electrical properties of water, particularly effective permittivity and conductivity, making capacitive sensing a promising method for concentration discrimination. Establishing a relationship between these parameters and the sensor response is essential for enabling real-time monitoring with aquatic robotic platforms. This section outlines the rationale for sensor selection and the design and fabrication of three custom interdigital sensor variants.

\subsection{Rationale for Interdigital Sensor Selection}

Among capacitive architectures, planar interdigital designs were chosen for their compact geometry, single-sided measurement capability, and ease of robotic integration. Unlike parallel-plate capacitors, which offer higher nominal capacitance but pose challenges for underwater use, interdigital sensors provide a planar structure that simplifies waterproofing and mounting on curved or confined surfaces. Their customizable electrode layout also allows for shaping the electric field distribution, benefiting embedded sensing applications.

However, planar designs typically exhibit lower capacitance and greater susceptibility to parasitic effects, especially inductance from long electrode fingers. To mitigate these, we limited digit length and employed impedance-based sensing at high frequencies. Low-frequency or DC capacitance measurement was avoided due to electrolysis risks and poor signal-to-noise ratios in the picofarad range.

To overcome limited absolute capacitance and improve discrimination, a VNA was used to characterize sensor impedance across a broad frequency sweep. This enabled identification of frequency bands where capacitive responses are both stable and sensitive to WAF concentration, informing robust sensing strategies for autonomous robotic deployment.

\subsection{Design Methodology and Fabrication}

The design of the three sensor variants ($x = 1, 2, 3$) was informed by: (1) geometric constraints of the test setup, and (2) the analytical model for interdigital capacitance \cite{bahl2022lumped}
\begin{equation}
\label{eq:capacitance}
    C \text{[pF]} = 10^{-3} \cdot \frac{\varepsilon_r}{18\pi} \cdot \frac{K(k)}{K'(k)} \cdot l \cdot(N-1),
\end{equation}
where $l$ is in microns, $N$ is the number of digits, and $\varepsilon_r$ is the effective dielectric constant. The ratio $\frac{K(k)}{K'(k)}$ accounts for fringing fields and depends on the geometry, using the parameters shown in Fig. \ref{fig:IDT}.

The sensors were designed to fit vertically into standard 25 mm × 150 mm test tubes, constraining the PCB footprint to 21.5 mm × 135 mm. All designs shared an active length $L = 105$ mm and gap $q = 0.5$ mm. A sensitivity analysis varying $N$, $w$, $s$, and $l$ by ±10\% showed $N$ and $l$ had the most significant influence on total capacitance and were prioritized for variation.

Using~\eqref{eq:capacitance} and geometric constraints, the relative capacitance of each design was estimated. Design 2, with 164 digits of 11{,}000~$\mu$m length and 140~$\mu$m spacing, had the highest capacitance. Design 3 followed closely, with 170 digits of 10{,}000~$\mu$m length and 425~$\mu$m spacing. Design 1, with 59 digits of 7{,}500~$\mu$m length and 850~$\mu$m spacing, had the lowest capacitance. As expected, longer, denser electrode layouts produced higher capacitance, suggesting greater sensitivity.

The sensors were fabricated on FR4 substrate using a CNC milling machine (ProtoMat S64, LPKF Laser \& Electronics, Germany). After milling, copper layers were cleaned with steel wool and isopropyl alcohol, then inspected for continuity. SMA connectors were soldered, and the boards were coated with a conformal acrylic layer (419D-340G, MG Chemicals Ltd, Canada) to insulate traces and prevent corrosion.

\section{Methods}
\label{sec:methods}

The performance of the fabricated interdigital sensors was evaluated through controlled laboratory experiments using WAFs with varying dilbit concentrations. A WAF represents the dissolved or dispersed components of a complex substance, such as oil, in water. Commonly used in toxicity testing and petroleum analysis, WAFs better reflect environmental exposure conditions than bulk oil \cite{aurand2005cooperative}. Given their reproducibility and ecological relevance, WAFs were selected to characterize the sensor response across known dilbit concentrations under repeatable conditions.

The methodology involved two components: (1) preparation of well-characterized WAF samples over a defined concentration range, and (2) acquisition of impedance data under consistent conditions using a VNA. Subsections~\ref{subsec:sample} and~\ref{subsec:experiments} detail these protocols.

\subsection{Sample Preparation}
\label{subsec:sample}

WAF preparation followed a small-volume protocol adapted from Adams et al. \cite{adams2020effects}, based on the standard published by Aurand and Coelho for the American Petroleum Institute \cite{aurand2005cooperative}. Fresh Cold Lake Blend dilbit, poured on February 3rd, 2025, was stored sealed and refrigerated until use on May 22nd, 2025.

Eleven dilutions, ranging from 0\% to 100\% WAF in 10\% increments, were prepared using distilled water and poured into labelled glass test tubes. Each tube was filled to 50 mL to maintain 20–30\% headspace, as recommended. The same guideline applied to the mixing container. All steps were performed in a fume hood due to dilbit's toxicity. Ambient temperature was monitored using a pre-calibrated DS18B20 sensor connected to an Arduino Uno with a real-time clock shield.

Glassware was rinsed, dried, and cleaned with isopropyl alcohol. For each capacitor design, the following protocol was used: 720~mL of distilled water and a 25~mm magnetic stir bar were added to a 1~L beaker and mixed at 220~rpm (adjusted to prevent vortex formation). After temperature stabilization, 80 mL of dilbit was gently added to the center. The beaker was covered with foil and stirred continuously for 18 hours to form the WAF.

To prepare dilutions, test tubes were pre-filled with specific volumes of distilled water (0–50~mL in 5~mL increments). The surface oil was gently displaced using a scoopula, and the aqueous phase was extracted using a 22-gauge gastight syringe (1000 Series, Hamilton Company, USA), then transferred to the corresponding test tube to complete the dilution.

\subsection{Experimental Setup and Methods}
\label{subsec:experiments}

\begin{figure}[t]
        \centering
    \subfloat[\label{fig:setup}]{%
       \includegraphics[width=0.52\linewidth]{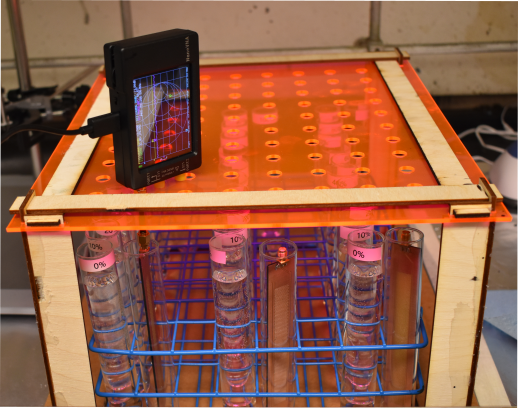}}
        \hfill
        \subfloat[\label{setup2}]{%
       \includegraphics[width=0.52\linewidth]{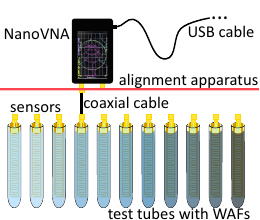}}
        \hfill
    \subfloat[\label{setup3}]{%
        \includegraphics[width=0.52\linewidth]{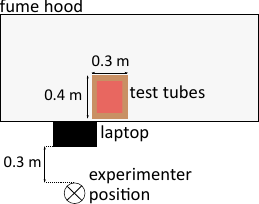}}
    \\
    \caption{Details of the experimental setup. (a) Test tube rack within the custom alignment apparatus, (b) frontal view diagram, and (c) top view diagram.}
    \label{fig:setup_details}
\end{figure}

Each filled test tube was randomly assigned one of the 33 fabricated sensors for impedance measurements. Before immersion, ten $S_{11}$ acquisitions were recorded in air as a reference for each sensor (Fig.~\ref{fig:setup}). During testing, sensors were fully immersed in the liquid samples, and measurements were collected using a handheld VNA (nanoVNA-H v1.2.27). The device was calibrated from 100~kHz to 300~MHz with 3201 data points, and ten acquisitions per sample were recorded using the NanoVNA-App (\href{https://nanovna.com/?page_id=141}{v1.1.208}, OneOfEleven).

The $S_{11}$ parameter, or input reflection coefficient, indicates the proportion of reflected power due to impedance mismatch. $S_{11} = 0$~dB corresponds to total reflection; more negative values (e.g., $-10$~dB) indicate better matching. The complex impedance $Z$ was extracted from $S_{11}$ using
\begin{equation}
Z = Z_0\frac{1 - \Gamma}{1 + \Gamma},
\end{equation}
where the reflection coefficient $\Gamma = \text{Re}(S_{11}) + j\text{Im}(S_{11})$, and $Z_0 = 50~\Omega$.

To reduce variability and ensure repeatability, some strategies were incorporated into the experimental setup and protocols. The same operator prepared all samples within four hours of WAF mixing, using consistent sources for WAF and distilled water. A custom fixture aligned each test tube relative to the VNA. The same coaxial cable (fixed length and orientation) was used throughout (Fig. \ref{setup2}). Sensor handling and positioning were consistent, and the same operator performed all acquisitions at the same workstation (Fig. \ref{setup3}).

\section{Results and Discussion} 
\label{sec:results}

This section presents the impedance characterization of the three sensor designs ($x = 1, 2, 3$) across dilbit WAF concentrations. The analysis, based on $S_{11}$-derived impedance, aims to inform sensor selection and operational strategy. Experimental temperature was held near \SI{18}{\celsius}.

\begin{figure}[b]
    \centering
    \includegraphics[width=\linewidth]{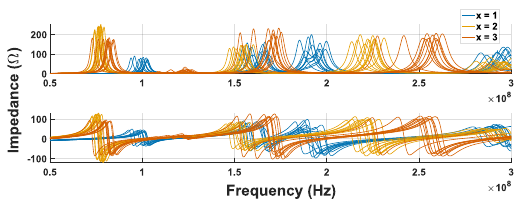}
    \caption{Impedance ($Z$) information recovered from the measured $S_{11}$ parameter for all evaluated concentrations by all capacitor designs in the full studied range. Top: real part of $Z$ and bottom: imaginary part of $Z$. }
    \label{fig:all}
\end{figure}

Due to measurement instability below 500~kHz, and lack of relevant variation up to 50~MHz, the analysis focuses on the 50–300~MHz range, where clearer and more consistent responses were observed (Fig.~\ref{fig:all}). All results reflect the average of ten repeated acquisitions per sample. Standard deviation was low throughout: median $\text{Re}(Z)$ deviations were below 0.014~$\Omega$, with maximums of 6.14, 2.90, and 7.35~$\Omega$ for sensors 1, 2, and 3. Variability in $\text{Im}(Z)$ was negligible.

\subsection{Interpreting Real and Imaginary Impedance Components}

\begin{figure}[t]
    \centering
    \includegraphics[width=\linewidth]{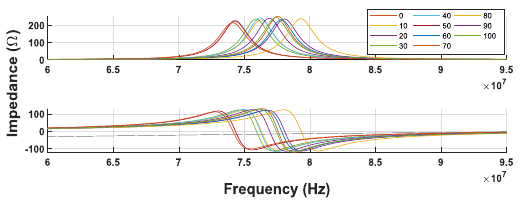}
    \caption{Impedance ($Z$) response of capacitor 2 between 60–95~MHz. Top: real part; bottom: imaginary part.}
    \label{fig:real}
\end{figure}

Fig.~\ref{fig:real} illustrates the impedance response of sensor 2 within a reduced frequency range for visualization. Similar trends were observed for sensors 1 and 3. In an ideal capacitive system, $\text{Re}(Z)$ would be negligible. However, due to the skin effect \cite{ramo1994fields}, resistance increases with frequency. While conductance (inverse of resistance) could serve as a sensing metric, it is generally less selective and more challenging to measure robustly in embedded systems. Capacitance-focused sensing offers better selectivity and integrates more easily with frequency-based readout circuits.

The $\text{Im}(Z)$ data reveal a frequency-dependent transition between capacitive (negative $\text{Im}(Z)$) and inductive (positive $\text{Im}(Z)$) behaviour. Inductive effects are dominant at higher frequencies and in long, narrow conductive structures. In underwater environments, nearby magnetic materials may amplify these effects. Therefore, analysis focuses on frequency bands where $\text{Im}(Z) < 0$, indicating a stable capacitive response suitable for concentration discrimination.

The optimal region occurs where vertical separation between impedance curves for different concentrations is maximized, improving resolution. However, operating near the resonance transition (where $\text{Im}(Z) = 0$) should be avoided due to instability. Notably, $\text{Re}(Z)$ tends to peak when $\text{Im}(Z)$ crosses zero, and these zero-crossing frequencies shift with concentration. This suggests they could serve as additional sensing features. In some cases, impedance values converge for different concentrations, reinforcing the need for restricted frequency bands or thresholding strategies in deployment.

\subsection{Capacitive Impedance Analysis and Best Frequency Selection}

To identify the most favourable frequencies, only those where $\text{Im}(Z) \leq 0$ across all concentrations were retained. Within these bands, the range of $\text{Im}(Z)$ values was computed as a function of frequency. Maximum variations were: 74.45~$\Omega$ at 199.80~MHz (sensor 1), 105.11~$\Omega$ at 79.39~MHz (sensor 2), and 104.30~$\Omega$ at 176.38~MHz (sensor 3).

For each capacitor design, all frequencies where the imaginary part of the impedance satisfied $\text{Im}(Z) \leq 0$ across all WAF concentrations were identified as valid candidates. At each of these frequencies, a score was computed using
\begin{equation}
\label{eq:score}
\text{Score}(f) = d_f + \alpha \cdot \Delta \text{Im}(Z)_f,
\end{equation}
where $d_f$ is the distance (in MHz) to the nearest $\text{Im}(Z)$ zero-crossing, $\Delta \text{Im}(Z)_f$ is the range of imaginary impedance across concentrations, and $\alpha = 0.001$ balances units (MHz vs. $\Omega$). The scoring function in \eqref{eq:score} balances two priorities: (i) stability of capacitive behaviour, quantified as the distance from the nearest resonance transition, and (ii) sensitivity to concentration, captured by the magnitude of impedance variation. The distance $d_f$ ensures operation away from unstable frequency regions where inductive and capacitive behaviour alternate rapidly. The impedance $\Delta \text{Im}(Z)_f$ quantifies sensitivity. This formulation prioritizes stable and sensitive operating points, without claiming physical dimensional homogeneity.

The frequency with the highest score was selected as the favourable operating point for each sensor design. The resulting selection is shown in Table \ref{tab:best_freq_range}. For capacitors 2 and 3, the frequencies obtained through this analysis were the same as where the impedance variation had already reached its maximum. 

\begin{table}[h]
\centering
\caption{Best Frequency and Corresponding Impedance Variation for Each Capacitor Design}
\label{tab:best_freq_range}
\begin{tabular}{c c c}
\hline
\textbf{$x$} & \textbf{Best Frequency (MHz)} & \textbf{Variation ($\Omega$)} \\
\hline
1 & 204.13 & 55.52 \\
2 & 79.39  & 105.11 \\
3 & 176.38 & 104.30 \\
\hline
\end{tabular}
\end{table}

The variation values reported in Table \ref{tab:best_freq_range} represent the range ($\Delta \text{Im}(Z)_f$) of imaginary impedance values across the eleven tested WAF concentrations. Higher variation indicates greater sensor response to concentration changes and is desirable for sensitivity. Capacitor 2 exhibited the most significant variation (105.11~$\Omega$ at 79.39~MHz), suggesting strong responsiveness to dielectric changes at that frequency. Capacitor 1 showed the lowest variation, consistent with its reduced capacitance and digit density. These outcomes align with design expectations and validate the use of $\text{Im}(Z)$ as a metric for sensor performance.  Variation may also be amplified at certain frequencies due to proximity to resonance transitions, which further emphasizes the importance of the scoring method in identifying stable operating points.

\begin{figure}[t]
    \centering
    \includegraphics[width=\linewidth]{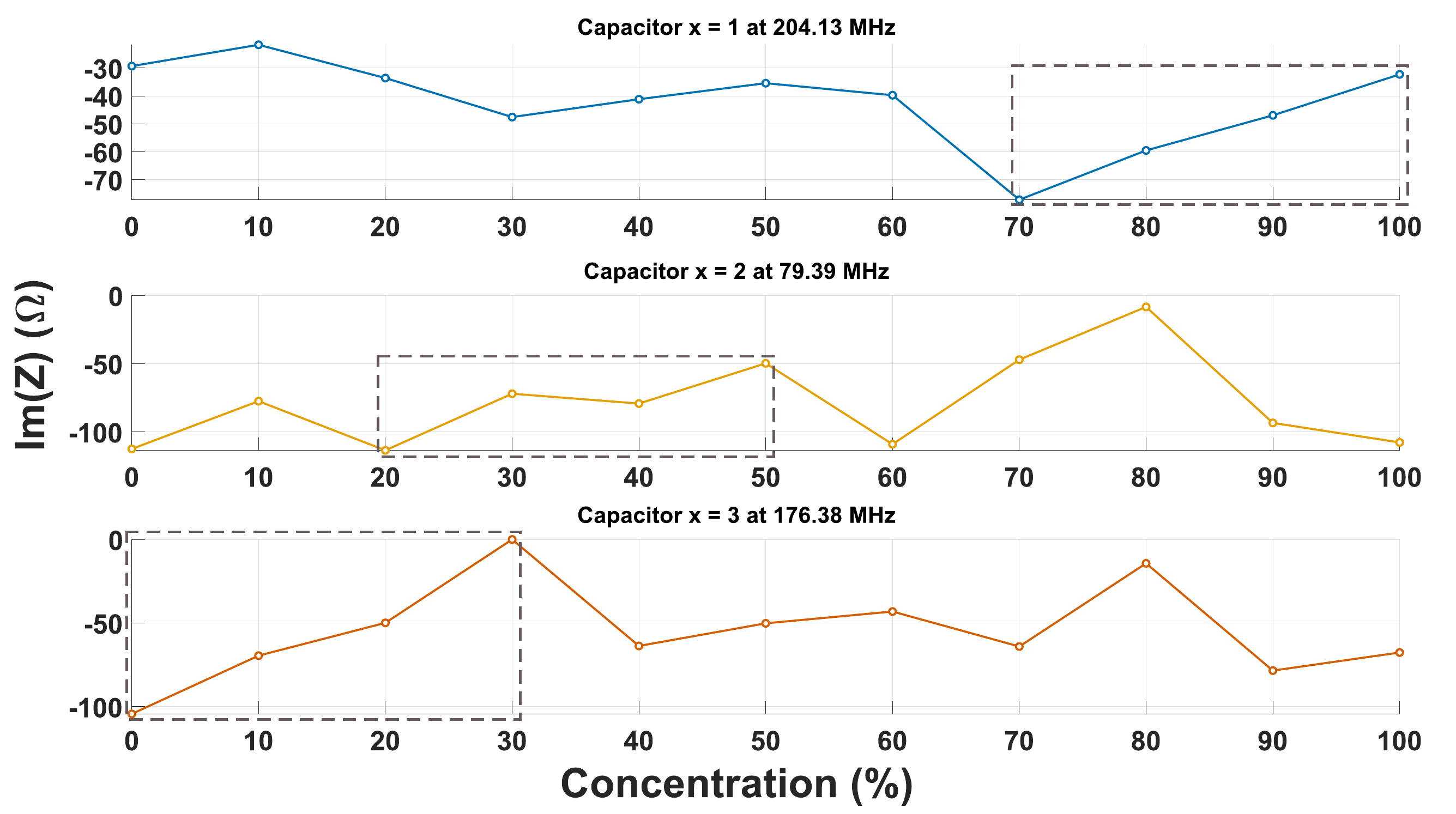}
    \caption{Imaginary impedance measurements for each concentration taken at the frequency where the capacitor displays the greatest sensitivity. From top to bottom: capacitor 1 at 204.13 MHz, capacitor 2 at 79.39 MHz, and capacitor 3 at 176.38 MHz. The dashed rectangles indicate regions where the sensor response is linear for four consecutive analyzed concentrations.}
    \label{fig:concentrations}
\end{figure}

Finally, the Im($Z$) values at the found capacitive frequencies were plotted as a function of concentration for each capacitor design, as shown in Fig. \ref{fig:concentrations}. These curves reflect the sensitivity of each design to changes in concentration and were used to assess the potential linearity of the sensor response. 

Analyzing this data reveals that not only do different dilbit concentrations alter key electrical properties of water, but also that capacitive sensors are sensitive to such changes, highlighting their potential in detecting and distinguishing between concentrations. Based on the impedance range data at optimal frequencies, the sensors demonstrated discernible responses to concentration differences as slight as 10\% WAF. This suggests that, even in their preliminary form, the sensors can reliably distinguish between modest concentration shifts, a critical capability for detecting the onset and spread of dilbit contamination.

Nonetheless, as seen in Table \ref{tab:best_freq_range}, capacitor designs $x$ = 2 and $x$ = 3 exhibited more prominent variation with concentration than $x$ = 1. However, each sensor design showcases a range of four analyzed concentrations where the behaviour is linear, as indicated by the dashed rectangles in Fig. \ref{fig:concentrations}. 
From a sensing perspective, linear behaviour is desirable due to simplified calibration, better measurement precision, and ease of predictability in system behaviour. 

Capacitor 2 demonstrated the highest variation in impedance across concentrations (105.11~$\Omega$) and maintained a clear linear region within a mid-range concentration window (20--50\% WAF). This makes it the most promising candidate for general-purpose sensing, especially in early spill detection scenarios. Capacitor 3, while nearly matching capacitor 2 in terms of impedance range (104.30~$\Omega$), exhibited linearity at lower concentrations, which could be beneficial for the detection of more diluted or dispersed dilbit. Capacitor 1, although exhibiting the lowest sensitivity and more limited linearity, showed consistent behaviour at higher concentrations, suggesting potential utility in detecting denser accumulations or proximity to the spill source.

These findings suggest that no single design is universally optimal. Instead, a multi-sensor approach, either as an array embedded in a single platform or distributed among specialized robotic agents, may enhance overall monitoring performance. Besides, for real-world deployment, environmental robustness must be considered. While the controlled lab conditions ensured stable measurements, factors such as temperature variation, ionic content, and water flow may affect impedance readings in situ. Future work should therefore include validation in simulated natural environments and robustness testing under expected operational conditions.

\subsection{Environmental Considerations}

Compared to fluorometry, which relies on the optical excitation and emission of polycyclic aromatic hydrocarbons and is less effective for weathered or sunken dilbit, the capacitive sensor approach offers the advantage of detecting electrical changes directly within the aqueous phase. Electrochemical sensors provide strong selectivity but often require complex integration and suffer from limited lifespan due to electrode fouling. In contrast, interdigital capacitive sensors are low-cost, robust, and planar. These are features that facilitate embedding in robotic agents for real-time, distributed sensing. While our approach trades off some chemical specificity for mechanical and electrical simplicity, this enables scalable deployment across fleets of aquatic robots.

Despite promising results under controlled laboratory conditions, several environmental challenges may impact sensor performance in real-world applications. Variations in water temperature can alter dielectric properties, while ionic content (e.g., from minerals or runoff) could introduce background impedance changes. Water turbulence and flow might affect sensor positioning and stability, and the presence of organic matter could gradually degrade sensor response.

This study used distilled water to isolate the effect of dilbit. While this ensured repeatability, it does not yet fully demonstrate selectivity. Future work will involve testing sensor response in the presence of common freshwater constituents and interferents to assess and improve specificity for dilbit detection.



\section{Conclusion} 
\label{sec:conclusion}

This study demonstrated the feasibility of using interdigital capacitive sensors for underwater detection of dilbit in freshwater environments. By analyzing the impedance response of three sensor designs across a range of WAF concentrations, we showed that variations in the medium’s electrical properties, particularly within the capacitive regime, can be effectively captured for concentration discrimination of fresh Cold Lake Blend dilbit. The approach combines mechanical simplicity with electrical sensitivity, supporting scalable deployment in swarm robotic systems.

By tuning sensor geometry, we found it possible to target distinct linear ranges at low, mid, or high concentrations. The complementary performance of each design suggests that multi-sensor or distributed deployments could improve both spatial and quantitative resolution for spill detection.

Impedance analysis prioritized the imaginary component ($\text{Im}(Z)$), which enhances selectivity and enables integration with frequency-based readout circuits, crucial for robotic applications. Such platforms could facilitate real-time, distributed monitoring of freshwater ecosystems, supporting early spill detection and delivering actionable data for environmental assessment and protection.

Future work will examine additional dilbit types and weathering conditions, conduct in situ validation, and improve environmental robustness. To address field deployment challenges, we will explore incorporating impedance models and thresholding strategies to filter non-dilbit-induced changes.


\section*{Acknowledgement}
We thank Dr. Bruce Hollebone (Environment and Climate Change Canada) for facilitating access to diluted bitumen and Dr. Julie Adams for guidance on sample preparation protocols. We are grateful to Diancheng Li for assistance with sensor manufacturing and to Dr. Xian Wang for providing access to the fume hood and essential laboratory resources. We also thank Dr. Elyson Carvalho for invaluable discussions and insights.

\bibliographystyle{IEEEtran}
\bibliography{bibliography}

\begin{thebibliography}{10}
\providecommand{\url}[1]{#1}
\csname url@samestyle\endcsname
\providecommand{\newblock}{\relax}
\providecommand{\bibinfo}[2]{#2}
\providecommand{\BIBentrySTDinterwordspacing}{\spaceskip=0pt\relax}
\providecommand{\BIBentryALTinterwordstretchfactor}{4}
\providecommand{\BIBentryALTinterwordspacing}{\spaceskip=\fontdimen2\font plus
\BIBentryALTinterwordstretchfactor\fontdimen3\font minus \fontdimen4\font\relax}
\providecommand{\BIBforeignlanguage}[2]{{%
\expandafter\ifx\csname l@#1\endcsname\relax
\typeout{** WARNING: IEEEtran.bst: No hyphenation pattern has been}%
\typeout{** loaded for the language `#1'. Using the pattern for}%
\typeout{** the default language instead.}%
\else
\language=\csname l@#1\endcsname
\fi
#2}}
\providecommand{\BIBdecl}{\relax}
\BIBdecl

\bibitem{stoyanovich2022fate}
S.~Stoyanovich, Z.~Yang, M.~Hanson, B.~Hollebone, D.~M. Orihel, V.~Palace, J.~Rodriguez-Gil, F.~Mirnaghi, K.~Shah, and J.~Blais, ``Fate of polycyclic aromatic compounds from diluted bitumen spilled into freshwater limnocorrals,'' \emph{Science of The Total Environment}, vol. 819, p. 151993, 2022.

\bibitem{rolsky2020municipal}
C.~Rolsky, V.~Kelkar, E.~Driver, and R.~U. Halden, ``Municipal sewage sludge as a source of microplastics in the environment,'' \emph{Current Opinion in Environmental Science \& Health}, vol.~14, pp. 16--22, 2020.

\bibitem{binding2021eolakewatch}
C.~Binding, L.~Pizzolato, and C.~Zeng, ``Eolakewatch; delivering a comprehensive suite of remote sensing algal bloom indices for enhanced monitoring of canadian eutrophic lakes,'' \emph{Ecological Indicators}, vol. 121, p. 106999, 2021.

\bibitem{su2021human}
G.~Su, M.~Logez, J.~Xu, S.~Tao, S.~Vill{\'e}ger, and S.~Brosse, ``Human impacts on global freshwater fish biodiversity,'' \emph{Science}, vol. 371, no. 6531, pp. 835--838, 2021.

\bibitem{national2016spills}
N.~A. of~Sciences, D.~on~Earth, L.~Studies, B.~on~Chemical~Sciences, and C.~on~the Effects of Diluted Bitumen on~the Environment, \emph{Spills of diluted bitumen from pipelines: A comparative study of environmental fate, effects, and response}.\hskip 1em plus 0.5em minus 0.4em\relax National Academies Press, 2016.

\bibitem{fingas2015diluted}
M.~Fingas, ``Diluted bitumen (dilbit): A future high risk spilled material,'' \emph{Proceedings of Interspill}, p.~24, 2015.

\bibitem{ji2024optimized}
W.~Ji, A.~G. Slaughter, G.~M. Coelho, T.~J. Nedwed, R.~C. Prince, and M.~C. Boufadel, ``Optimized underwater detection of dispersed oils using scanning fluorometry,'' in \emph{International Oil Spill Conference Proceedings}, vol. 2024, no.~1.\hskip 1em plus 0.5em minus 0.4em\relax Allen Press, 2024.

\bibitem{dunbabin2012robots}
M.~Dunbabin and L.~Marques, ``Robots for environmental monitoring: Significant advancements and applications,'' \emph{IEEE Robotics \& Automation Magazine}, vol.~19, no.~1, pp. 24--39, 2012.

\bibitem{bogue2023role}
R.~Bogue, ``The role of robots in environmental monitoring,'' \emph{Industrial Robot: the international journal of robotics research and application}, vol.~50, no.~3, pp. 369--375, 2023.

\bibitem{lindsay2022collaboration}
J.~Lindsay, J.~Ross, M.~L. Seto, E.~Gregson, A.~Moore, J.~Patel, and R.~Bauer, ``Collaboration of heterogeneous marine robots toward multidomain sensing and situational awareness on partially submerged targets,'' \emph{IEEE Journal of Oceanic Engineering}, vol.~47, no.~4, pp. 880--894, 2022.

\bibitem{bedenik2025bistable}
G.~Bedenik, A.~Morales, S.~Pieris, B.~da~Silva, J.~W. Kurelek, M.~Greeff, and M.~Robertson, ``Bistable sma-driven engine for pulse-jet locomotion in soft aquatic robots,'' \emph{arXiv preprint arXiv:2504.03988}, 2025.

\bibitem{haghighi2025innovative}
S.~K. Haghighi, S.~Mohammadlou, S.~Angizi, and A.~Hatamie, ``Innovative electrochemical nano-robot: Integrating printed nanoelectronics with a remote-controlled robotic for on-site underwater electroanalysis,'' \emph{Langmuir: the ACS journal of surfaces and colloids}, vol.~41, no.~13, pp. 8592--8601, 2025.

\bibitem{saleh2018situ}
M.~Saleh, G.~Oueidat, I.~H. Elhajj, and D.~Asmar, ``In situ measurement of oil slick thickness,'' \emph{IEEE Transactions on Instrumentation and Measurement}, vol.~68, no.~7, pp. 2635--2647, 2018.

\bibitem{saleh2022dual}
M.~Saleh, A.~R. Tabikh, I.~H. Elhajj, K.~McKinney, and D.~Asmar, ``Dual-modality capacitive-ultrasonic sensing for measuring floating oil spill thickness,'' \emph{IEEE Transactions on Instrumentation and Measurement}, vol.~71, pp. 1--14, 2022.

\bibitem{saleh2024situ}
M.~Saleh, I.~H. Elhajj, and D.~Asmar, ``In situ sensors for oil spill detection and thickness measurement: Methods and challenges,'' \emph{IEEE Transactions on Instrumentation and Measurement}, 2024.

\bibitem{teodorescu2024constraints}
H.-N.~L. Teodorescu, ``Constraints on planar and planar interdigital capacitors used in sensors,'' in \emph{2024 16th International Conference on Electronics, Computers and Artificial Intelligence (ECAI)}.\hskip 1em plus 0.5em minus 0.4em\relax IEEE, 2024, pp. 1--6.

\bibitem{bahl2022lumped}
I.~J. Bahl, \emph{Lumped elements for RF and microwave circuits}.\hskip 1em plus 0.5em minus 0.4em\relax Artech house, 2022.

\bibitem{aurand2005cooperative}
D.~Aurand and G.~Coelho, ``Cooperative aquatic toxicity testing of dispersed oil and the chemical response to oil spills: Ecological effects research forum (croserf),'' \emph{Inc. Lusby, MD. Tech. Report}, pp. 07--03, 2005.

\bibitem{adams2020effects}
J.~E. Adams, B.~N. Madison, K.~Charbonneau, M.~Sereneo, L.~Baillon, V.~S. Langlois, R.~S. Brown, and P.~V. Hodson, ``Effects on trout alevins of chronic exposures to chemically dispersed access western blend and cold lake blend diluted bitumens,'' \emph{Environmental Toxicology and Chemistry}, vol.~39, no.~8, pp. 1620--1633, 2020.

\bibitem{ramo1994fields}
S.~Ramo, J.~R. Whinnery, and T.~Van~Duzer, \emph{Fields and waves in communication electronics}.\hskip 1em plus 0.5em minus 0.4em\relax John Wiley \& Sons, 1994.

\end{thebibliography}

\end{document}